\RequirePackage{lineno}
\documentclass[twocolumn,showpacs,superscriptaddress,amsmath,amssymb]{revtex4-1}
\usepackage{graphicx}
\usepackage{epsfig}
\usepackage{dcolumn}
\usepackage{bm}
\usepackage{overpic}
\usepackage{color}
\usepackage{ulem}
\usepackage{lineno}
\usepackage{subfigure}

\newcommand{\ra}{\rightarrow}
\newcommand{\lamlam}{e^{+}e^{-}\ra\Lambda\bar{\Lambda}}
\newcommand{\lambar}{\bar{\Lambda}}
\newcommand{\lam}{\Lambda}
\newcommand{\lamdec}{\Lambda\ra p\pi^{-}}
\newcommand{\lambardec}{\bar{\Lambda}\ra\bar{p}\pi^{+}}
\newcommand{\lambarneu}{\bar{\Lambda}\ra\bar{n}\pi^{0}}
\newcommand{\ee}{e^{+}e^{-}}
\newcommand{\ppi}{p\pi^{-}}
\newcommand{\pbarpi}{\bar{p}\pi^{+}}
\newcommand{\npi}{\bar{n}\pi^{0}}
\newcommand{\pppipi}{p\bar{p}\pi^{+}\pi^{-}}
\newcommand{\sqs}{\sqrt{s}}
\begin{document}
\title{\boldmath
Observation of a cross-section enhancement near mass threshold in $e^{+}e^{-}\to\Lambda\bar{\Lambda}$
}
\author{
  \small
M.~Ablikim$^{1}$, M.~N.~Achasov$^{9,d}$, S. ~Ahmed$^{14}$, X.~C.~Ai$^{1}$, O.~Albayrak$^{5}$, M.~Albrecht$^{4}$, D.~J.~Ambrose$^{45}$, A.~Amoroso$^{50A,50C}$, F.~F.~An$^{1}$, Q.~An$^{47,38}$
% ...
\\
 \vspace{0.2cm}
 (BESIII Collaboration)\\
 \vspace{0.2cm} 
{\it
$^{1}$ Institute of High Energy Physics, Beijing 100049, People's Republic of China\\
% ...
$^{4}$ Bochum Ruhr-University, D-44780 Bochum, Germany\\
$^{5}$ Carnegie Mellon University, Pittsburgh, Pennsylvania 15213, USA\\
% ...
$^{9}$ G.I. Budker Institute of Nuclear Physics SB RAS (BINP), Novosibirsk 630090, Russia\\
% ...
$^{14}$ Helmholtz Institute Mainz, Johann-Joachim-Becher-Weg 45, D-55099 Mainz, Germany\\
% ...
$^{38}$ State Key Laboratory of Particle Detection and Electronics, Beijing 100049, Hefei 230026, People's Republic of China\\
% ...
$^{45}$ University of Rochester, Rochester, New York 14627, USA\\
% ...
$^{47}$ University of Science and Technology of China, Hefei 230026, People's Republic of China\\
% ...
$^{50}$ (A)University of Turin, I-10125, Turin, Italy; (B)University of Eastern Piedmont, I-15121, Alessandria, Italy; (C)INFN, I-10125, Turin, Italy\\
% ...
$^{d}$ Also at the Novosibirsk State University, Novosibirsk, 630090, Russia\\
% ...
 }
\vspace{0.4cm}
}

\date{\today}

\begin{abstract}
The process $\lamlam$ is studied using data samples at $\sqrt{s}=2.2324$, 2.400, 2.800 and 3.080~GeV collected with the BESIII detector
operating at the BEPCII collider.
The Born cross section is measured at $\sqrt{s}$=2.2324~GeV, which is 1.0~MeV
above the $\lam\lambar$ mass threshold, to be $305\pm45^{+66}_{-36}$~pb, where the first uncertainty is statistical and the second systematic.
The cross section near threshold
is larger than that expected from theory, which predicts the cross section to vanish at threshold.
The Born cross sections at $\sqrt{s}$=2.400, 2.800 and 3.080~GeV are
measured and found to be consistent
with previous experimental results, but with improved precision.
Finally, the corresponding effective electromagnetic form factors of $\lam$ are deduced.
\end{abstract}

\maketitle

\section{\boldmath INTRODUCTION}
Electromagnetic form factors~(FFs) are important observables
for probing the inner structure of
hadrons and for understanding the strong interaction.
The time-like FFs are mostly measured by electron-positron colliding experiments~\cite{ee}.
The Born cross section for the process $\ee\ra B\bar{B}$ via one-photon exchange,
where $B$ is a spin 1/2 baryon, can be expressed in terms of the electric
and magnetic FFs $G_{E}$ and $G_{M}$,
\begin{equation}
\label{equ:eq1}
 \sigma^{B}(s)=\frac{4\pi\alpha^{2} C \beta}{3s}\left[|G_{M}(s)|^{2}+\frac{2m_{B}^{2}c^{4}}{s}|G_{E}(s)|^{2}\right].
\end{equation}
Here, $\alpha=1/137.036$ is the fine-structure constant,
$\beta=\sqrt{1-4m_{B}^{2}c^{4}/s}$ is the velocity, $c$ is the speed
of light, $s$ is the square of the center-of-mass~(c.m.) energy,
and $m_{B}$ is the mass of the baryon.
The Coulomb correction factor $C$~\cite{schwinger,coulomb},
accounting for the electromagnetic interaction of charged point-like fermion pairs in the final state,
is 1.0 for pairs of neutral baryons and
$y/(1-e^{-y})$ with $y=\pi\alpha(1+\beta^{2})/\beta$
for pairs of charged baryons.
The effective FF defined by
\begin{equation}
\label{eq2}
|G|\equiv\sqrt{\frac{|G_{M}(s)|^{2}+(2m_{B}^{2}c^{4}/s)\,|G_{E}(s)|^{2}}{1+2m_{B}^{2}c^{4}/s}}
\end{equation}
is proportional to the square root of the baryon pair production cross section.

Experimentally, there have been many  studies on the nucleon pair production cross sections and the time-like nucleon FFs in the
past decades~\cite{DM1,DM2,FENICE,cleoform,babar2,bes3,ps170, E760,E835,SND}.
Unusual behavior in the near-threshold region has been observed for both
$e^{+}e^{-}\rightarrow p\bar{p}$ and $e^{+}e^{-}\rightarrow n\bar{n}$ cross sections~\cite{babar2,ps170,SND}.
Compared with neutrons, the production cross section and FFs of
hyperons are however hardly explored~\cite{dm2,Babar,CLEOC}.
The BaBar experiment measured the hyperon final states of $\lam\lambar$~\cite{Babar} with significantly larger uncertainties
compared to the proton case.
The cross section of $\lamlam$ in a wide c.m.\ energy region
from threshold to $\sqrt{s}=2.27$~GeV was measured to be $204\pm60\pm20$~pb, which
indicates a possible non-vanishing cross section at threshold.
Recently, the BESIII experiment has observed a non-zero cross section near
the $\Lambda_{c}^{+}\bar{\Lambda}_{c}^{-}$ production threshold in the process
$e^{+}e^{-}\to\Lambda_{c}^{+}\bar{\Lambda}_{c}^{-}$~\cite{lambdac}.
The unexpected features of baryon pair production near threshold have driven many
theoretical interests~\cite{theory}, including scenarios that invoke $B\bar{B}$ bound
states or unobserved meson resonances.
It was also interpreted as an attractive Coulomb interaction on the constituent quark level in Ref.~\cite{cite:rinaldo}.
In order to properly test the hypotheses, a precision measurement of $\ee\ra \lam\lambar$
very close to $\lam\lambar$ mass threshold is needed.

In this paper, we present a study of process $\lamlam$ at c.m.~energy $\sqs=2.2324$~GeV, which is 1.0~MeV above the $\lam\lambar$ mass threshold,
with two decay modes reconstructed, $\lamdec$, $\lambardec$ (referred
to as mode I) and $\lambarneu$, $\Lambda\to X$ (referred to as mode II, where $X$ represents the inclusive decay of $\lam$).
Besides, measurements on the process $\lamlam$  at c.m.~energies $\sqrt{s}=2.400$, 2.800, and 3.080~GeV
are given with improved precision compared with previous experiments.

\section{\boldmath THE BESIII EXPERIMENT AND THE DATA SETS}
The collision data were taken with the BESIII spectrometer at BEPCII.
BEPCII is a double-ring $e^{+}e^{-}$ collider running at c.m.~energies between 2.0-4.6 GeV and reaches a peak luminosity of $1.0\times10^{33}$cm$^{-2}$s$^{-1}$ at a c.m.~energy of 3770 MeV.
The cylindrical BESIII detector has an effective geometrical acceptance of $93\%$ of 4$\pi$ and divides into a barrel section and two endcaps.
It contains a small cell, helium-based (60$\%$ He, 40$\%$ C$_{3}$H$_{8}$) main drift chamber (MDC) which provides momentum measurement of charged particle with a resolution of $0.5\%$ at a momentum of 1 GeV/c in a magnetic field of 1 Tesla.
The energy loss measurement ($dE/dx$) provided by the MDC has a resolution better than $6\%$.
A time-of-flight system (TOF) consisting of 5-cm-thick plastic scintillators can measure 
the flight time of charged particles with a time resolution of 80 ps in the barrel and 110 ps in the end-caps.
An electromagnetic calorimeter (EMC) consisting of 6240 CsI (Tl) in a cylindrical structure and two end-caps 
is used to measure the energies of photons and electrons.
The energy resolution of the EMC is $2.5\%$ in the barrel and $5.0\%$ in the end-caps for photon/electron of 1 GeV energy.
The position resolution of the EMC is 6 mm in the barrel and 9 mm in the end caps.
A detailed description of the detector and its performance can be found in Ref.~\cite{BESIII}.

A {\sc Geant4}-based~\cite{geant4} Monte Carlo~(MC) simulation software package, {\sc Boost}~\cite{boost} is used to generate the signal and background MC samples.
The signal process of $\lamlam$ at $\sqs=2.2324$~GeV is generated uniformly in phase space ({\sc PHSP}) since
$G_{E}$ equals to $G_{M}$ at threshold by definition~\cite{bes3}.
The corresponding correction factor is calculated by taking
the higher-order processes with one radiative photon in the final states and
the energy spread of collider beams into consideration, where the energy
spread is inversely proportional to the beam energy, to be 0.48~MeV from
a scaling at $J/\psi$ peak.
The subsequent decays of $\lamdec$, $\lambardec$ for mode I,
and $\lambarneu$, $\lam\to X$ for mode II are
generated with {\sc EvtGen}~\cite{evtgen}.
The signal process of $\lamlam$ at $\sqs=2.400$, 2.800 and 3.080~GeV is generated with
the software package {\sc Conexc}~\cite{conexc}, which includes correction factors
for higher-order processes with one radiative photon.
Simulated samples of the QED background processes $\ee\ra l^{+}l^{-}~(l=e,~\mu)$
and $\ee\ra\gamma\gamma$ are generated with {\sc Babayaga}~\cite{BABAYAGA}.
The generic (`inclusive') MC samples for hadronic final states from $\ee$ collision are generated with {\sc LundArea}~\cite{LUND}.

\section{\boldmath Reconstruction of $\lamlam$ at $\sqrt{s}$=2.2324~GeV}
The process $\lamlam$ at $\sqs$=2.2324~GeV
is selected via two decay modes, with the final state
topologies $\pppipi$ and $\npi X$.
Due to the near-threshold production and small PHSP
in $\lam(\lambar)$ decays, the nucleon and antinucleon
in the final state are difficult to detect.
Thus it is impossible to fully reconstruct the final states.
Instead, we employ an indirect search for the antiproton in mode I and
search for mono-energetic $\pi^{0}$ in mode II, respectively.

For mode I, the low momentum pions from signal final states can be detected directly.
A good charged track must have
a polar angle $\theta$ within $|\cos\theta|<0.93$ and have a point
of closest approach to the interaction point (IP) in the plane perpendicular to the beam, $V_{xy}$, within 1~cm and
along the beam direction, $V_{z}$, within $10$~cm.
The combined information of specific ionisation ($dE/dx$) and the time-of-flight (TOF) system is used to calculate particle identification (PID) probabilities
for the pion, kaon and proton hypotheses,
and the particle type with the highest probability is
assigned to the track. The candidate events are required to have two good charged tracks
identified as one positive and one negative pion,
and the momentum of the charged pions is required to be within [0.08, 0.11]~GeV/$c$,
as expected from $\lam(\lambar)$ decay.

\begin{figure}[htbp]
\begin{center}
\begin{overpic}[width=4.25cm,height=3.7cm,angle=0]{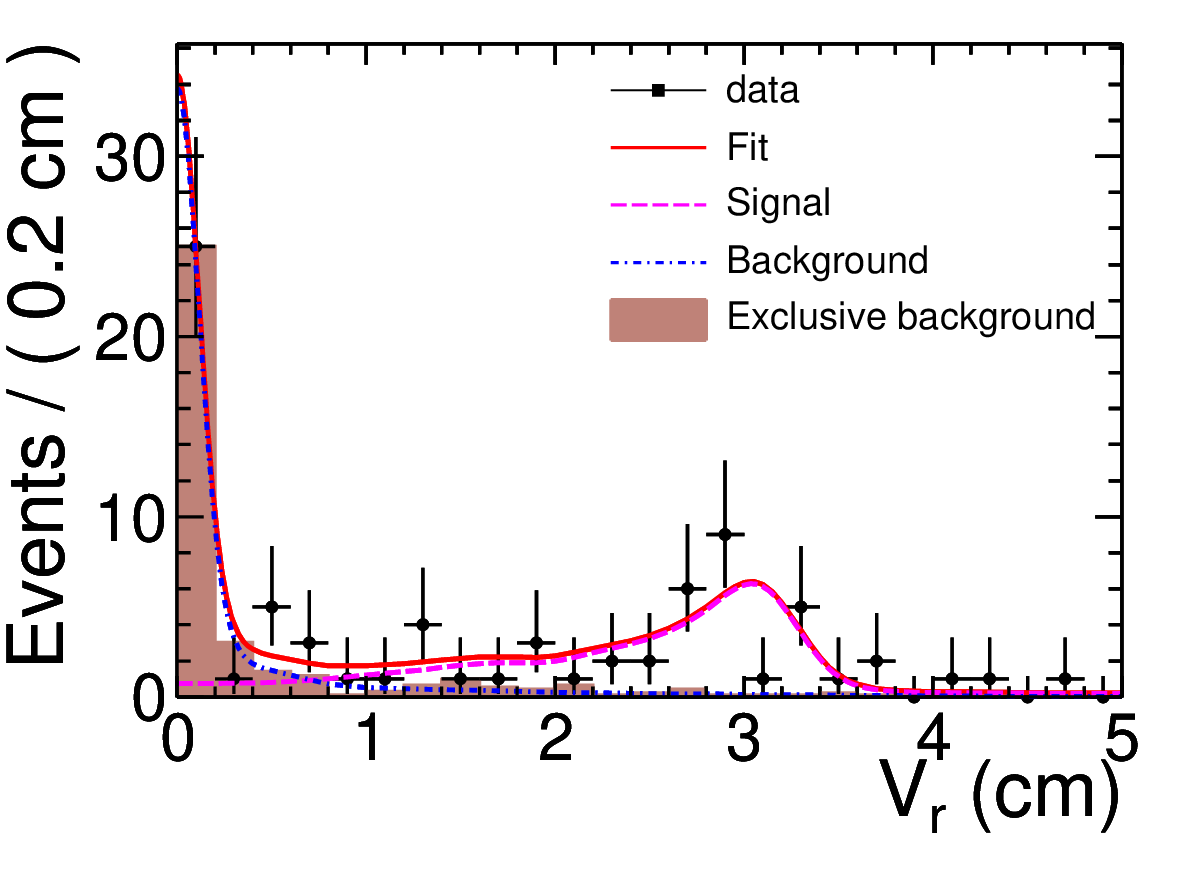}
\put(30,65){(a)}
\end{overpic}
\begin{overpic}[width=4.25cm,height=3.7cm,angle=0]{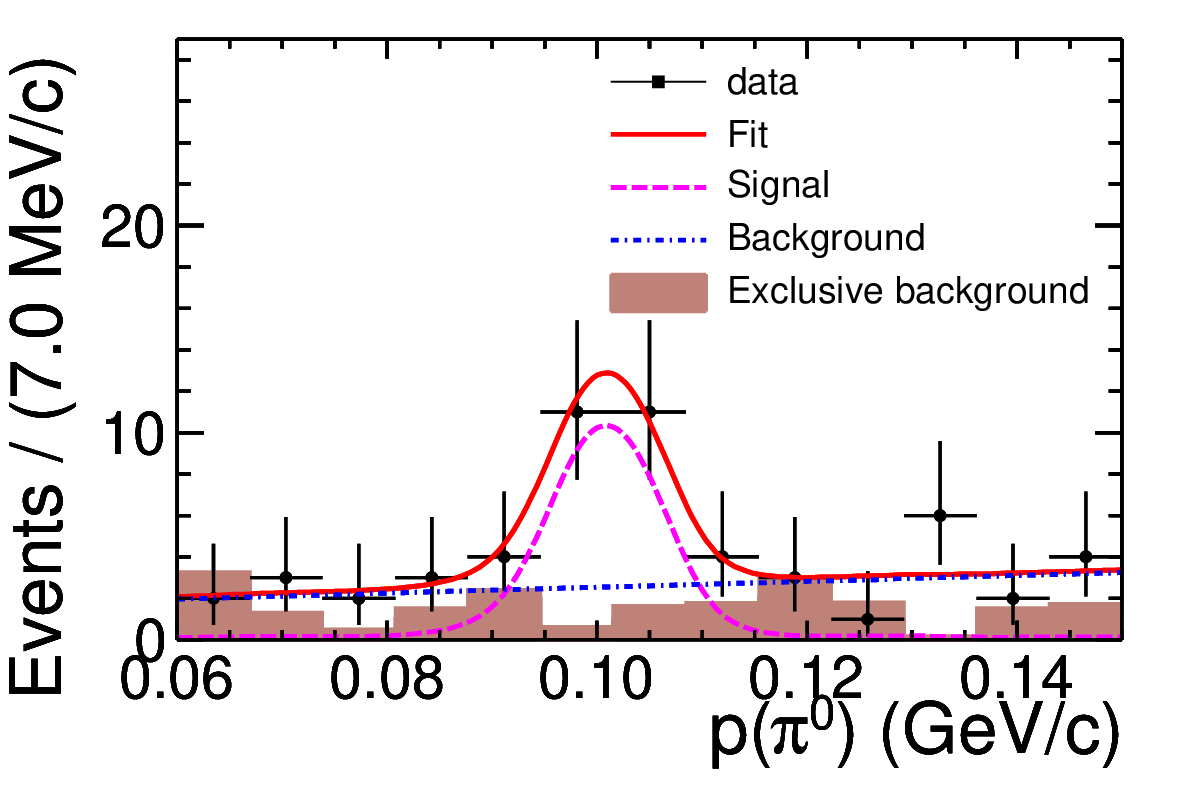}
\put(30,65){(b)}
\end{overpic}
\setlength{\abovecaptionskip}{-0.1cm}
\setlength{\belowcaptionskip}{-0.7cm}
\caption{(a) Fitted $V_{r}$ distribution for mode I and (b) fitted $p(\pi^{0})$ for mode II.
The dots with error bars are data,
the solid curves (red) are the fit results,
the dashed curves (pink) show the signals,
the dash-dot curves (blue) show the backgrounds
and the shaded histograms are the summed background from exclusive background process
which mainly stem from hadronic final states for mode I, hadronic final states and beam-associated backgroun}
\label{fitvr}
\end{center}
\end{figure}

The antiproton annihilates from signal final states in the interaction with nucleons of detector materials,
mostly in the beam pipe, and produces secondary particles.
The distribution of the largest $V_{xy}$ of all tracks apart from the two good charged pions in an
event, $V_{r}$, as shown in Fig.~\ref{fitvr}(a),
shows an enhancement around 3~cm, which is the
distance from the IP to the beam pipe.
After applying the above selection criteria,
the inclusive background processes in our MC data sample
do not contribute to the enhancement.
Besides, based on a study of pion momentum sidebands, which
are the events located in $[0.15, 0.18]$~GeV$/c$ of pion
momentum, there is no peaking background around 3~cm in the $V_{r}$
distribution.
The number of signal events for mode I is extracted by fitting the $V_{r}$ distribution, where the signal is described by the MC shape, the background is described by the sideband of the $\pi$ momentum as it is consistent with the distribution from inclusive background processes.

For mode II, at most one good charged track is allowed
and at least three neutral candidates are required.
Neutral candidates are required to have a minimum energy of 25~MeV in the barrel region
or 50~MeV in the endcap region.
To eliminate showers produced by charged particles,
the neutral candidates are required to
have no associated charged tracks within $10$ degrees.
The most energetic shower
is assumed to be a $\bar{n}$ and others to be photons,
motivated by MC simulations,
since the $\bar{n}$ annihilates with material in electromagnetic
calorimeter~(EMC) and produces several secondary particles with total
energy up to $2m_{n}$~GeV, where $m_{n}$ is the mass of neutron from the Particle Data Group~(PDG)~\cite{PDG}.
A $\pi^{0}$ candidate is identified by
a one-constrained~(1C) kinematic fit on $\pi^{0}$ mass
applied to each photon pair.
The energy asymmetry $|E_{\gamma_{1}}-E_{\gamma_{2}}|/p_{\pi^{0}}$ is required to be less than $0.95c$.
Furthermore, the angle between the momentum directions of the $\pi^{0}$ and $\bar{n}$ candidates is required to be larger
than $140^{\circ}$.
If there are several photon pair combinations, the one giving the smallest $\chi^{2}_{1C}$
is identified as the $\pi^{0}$ candidate.
To improve the signal-to-background ratio, only events with $\chi^{2}_{1C}<20$ are accepted.

After the preliminary selection for mode II, most of background events from QED processes can be removed. The inclusive hadronic final states with multiple $\pi^{0}$s and the beam-associated background events~\cite{touchek}
are the dominant background sources.
A dedicated data sample collected with BESIII with non-colliding beams is used to study the beam-associated
background and is described in Ref.~\cite{bes3}.
To separate the signal from the background, mainly to distinguish between $\bar{n}$ and $\gamma$, the boosted decision tree (BDT)
technique~\cite{BDT} is used in this analysis.
The signal events are generated with a {\sc PHSP} generator.
The background events are mixtures of hadronic final states and separated-beam events, where the
number of hadronic
final states is normalized according to the  luminosity, and the number of separated-beam events is normalized
to the remaining number of events in data.
Separate sets of BDTs are built with eight discriminating variables.
All input variables are EMC-related and are shown in Table~\ref{var},
together with their normalized importance values.
An optimal classifier requirement is applied for the BDT output.
The distribution of $\pi^{0}$ momentum in data and remaining background after the full selection is shown in Fig.~\ref{fitvr}(b),
where a clear enhancement can be observed around 0.1~GeV/$c$.
An unbinned maximum likelihood fit is performed to determine the number of
signal events, where the signal is described by the
MC shape convoluted with a Gaussian function, and the background is described by
a linear function which appears to give a good description for the background.
\begin{table}[htbp]
\caption{The variables used in the BDT classifier, ranked by the importance. }\label{var}
\begin{center}
\footnotesize
\begin{tabular}{c|c|c}
\hline
\hline
~Rank~~~ & ~~~~~~~~~~Variable~~~~~~~~~ & ~~~Importance~~\\
\hline
1  & energy deposition within 40$^{\circ}$ cone & $2.4\times10^{-1}$ \\
2  & deposited energy    & $2.0\times10^{-1}$  \\
3  & deposit of energy seed   &  $1.3\times10^{-1}$  \\
4  & Num. of hits within 40$^{\circ}$ cone &  $1.1\times10^{-1}$  \\
5  & Num. of hits &  $1.0\times10^{-1}$\\
6  & lateral moment   &  $9.3\times10^{-2}$  \\
7  & second moment  &  $7.6\times10^{-2}$  \\
8  & deposition shape~\cite{shape} &  $5.4\times10^{-2}$   \\
\hline
\hline
\end{tabular}
\end{center}
\end{table}

The Born cross section of the process $e^{+}e^{-}\to\lam\lambar$ is determined from
\begin{equation}
\sigma^{\rm B} = \frac{N_{\rm obs}}
            {\mathcal{L}_{\rm int}~\epsilon(1+ \delta)~\mathcal{B}},
\label{equ:born}
\end{equation}
where $\mathcal{L}_{\rm int}$ is the integrated luminosity,
$\epsilon$ is
the detection efficiency obtained from the signal MC sample, and
$1+\delta$ is the radiative correction factor, determined
by taking the energy spread and
initial state radiation (ISR)
photon emission
and the vacuum polarization into account.
$\mathcal{B}$ is the product of decay branching fractions of intermediate states
$\lamdec$ and $\lambardec$ for mode I,
$\lambarneu$, $\lam\to X$ and $\pi^{0}\ra \gamma\gamma$ for mode II.

Several sources of systematic uncertainties
are considered for the determination of the cross section at $\sqs=2.2324$~GeV.
The uncertainties for reconstruction of low momentum pions
for mode I are studied with $J/\psi\ra\pppipi$.
The differences of the efficiencies between data and MC are taken as the uncertainties,which give 12.3\% for tracking and 1.0\% for PID.
The uncertainty of the $V_{r}$ selection in mode I is also studied from the $J/\psi\ra\pppipi$ process.
The recoil momentum of $p\pi^{+}\pi^{-}$ is required to be within [0.08, 0.12]~GeV/$c$ to match the kinematics of the $\lambar$ decay.
The efficiency is defined as the number of events with $V_{r}$ less than 5~cm to the number of events with the recoil
mass of $p\pi^{+}\pi^{-}$ lying in the antiproton mass region. This uncertainty is estimated to be $0.3\%$.
The uncertainties of $\bar{n}$ and $\pi^{0}$ selection for mode II are studied in the processes,
to be 2.2\% and 2.3\% respectively.
The uncertainty associated with the BDT output requirement for mode II is estimated by selecting $\bar{n}$ sample
from process $J/\psi\ra p\bar{n}\pi^{-}$. The efficiency is obtained by applying the same classifier on
data and MC, and the difference
of 4.8\% is taken as the uncertainty.
The uncertainty due to the fit procedure is investigated by replacing the background shape with
an inclusive MC distribution and varying the fit range,
and the uncertainties are found to be 4.6\% and 8.8\% for modes I and II.
The uncertainty of the ISR correction is studied
by changing the cross section line shape of $e^{+}e^{-}\to\Lambda\overline{\Lambda}$ in the MC generator and then taking the difference of
$^{+18.5}_{-3.6}$\%
in the obtained ISR correction factor as the uncertainty.
The uncertainty due to the energy spread correction is 2.0\% by taking an alternative energy spread value
from another $\psi(3686)$ scan.
From a measurement of $J/\psi$ meson parameters~\cite{xingyu},
there is a nominal energy measurement uncertainty, 0.59~MeV, by comparing the mass of
reconstructed $J/\psi$ meson with the mass from PDG.
Since the two data sets are collected in the same data-taking period, we treat the
uncertainty of energy measurement in this analysis to be the same as $J/\psi$. 
By interpolating the line-shape with c.m.~energy value, the difference on $\epsilon (1+\delta)$, $3.9\%$,
is taken as the uncertainty from nominal c.m.~energy measurement.
The uncertainty of the trigger efficiency is 1.0\% for mode II~\cite{trigger}.
The uncertainty of the integral luminosity is 1.0\%, as determined from large-angle Bhabha events~\cite{lumi}.
Assuming all the sources of systematic are independent, the total uncertainties
are obtained by adding the individual contributions in quadrature, to be $^{+23.2}_{-14.4}$\% and $^{+22.1}_{-12.6}$\%
for modes I and II, respectively.

The systematic uncertainty in the effective FF $|G|$ can be derived from Eq.~(\ref{eq2}). It is half of that of
the Born cross section for the uncertainty sources not related with c.m~energy.
For the uncertainty from nominal c.m.~energy measurement and  energy spread,
due to the rapid variation of the velocity $\beta$ versus the c.m.~energy
near threshold,
large uncertainties are taken into consideration, to be $^{+22.6}_{-9.3}\%$ for energy shift
and $^{+15.2}_{-9.7}\%$ for energy spread, respectively.

\section{\boldmath Reconstruction of $\lamlam$ at other energy points}
The analysis at c.m.~energies $\sqs=2.400$, 2.800 and 3.080~GeV is straightforward
since full reconstruction of the final state $\pppipi$ is feasible.
Four good charged tracks with $V_{xy}$ within 10~cm and
$V_{z}$ within $30$~cm, identified as one proton-antiproton pair and one
pion pair $(\pi^{+}\pi^{-})$ are required.
Candidates for $\lam (\lambar)$ are reconstructed with proton and pion tracks.
A secondary vertex fit is performed and the track parameters are used to obtain the invariant
mass $M_{\ppi} (M_{\pbarpi})$. The mass window requirement $|M_{p\pi}-M_{\lam}|<0.01$~GeV/$c^{2}$
is used to select $\Lambda$($\overline{\Lambda}$) candidates,
where $M_{\lam}$ is the nominal mass of $\lam$ from the PDG~\cite{PDG}.
Further, c.m.\ energy dependent, requirements on the opening angle between $\lam$ and $\lambar$
in the center-of-mass system, $\theta_{\lam\lambar}>170^{\circ}$, $176^{\circ}$, $178^{\circ}$ at
$\sqrt{s}=2.400$, 2.800, 3.080~GeV are applied.

The background of the $\lamlam$ channel
either comes from non-$\lam$ background or $\lam$ peaking background.
The non-$\Lambda$ background is studied from the two-dimensional sideband of $M_{\ppi}$ and
$M_{\pbarpi}$.
The sideband regions $1.084<M_{p\pi^{+}/\bar{p}\pi^{+}}<1.104$ GeV/$c^{2}$
are defined to investigate the potential background without $\lam$ or $\lambar$ in the final states.
The $\lam$ peaking background is studied from the exclusive processes,
$\ee\ra\Sigma^{0}\bar{\Sigma}^{0}$,
$\ee\rightarrow\lam\bar{\Sigma}^{0}$ and $\ee\ra \Xi^{0}\bar{\Xi}^{0}$.
After applying the same selection criteria for the MC samples of these background channels
with luminosity normalized,
the numbers of surviving background events
are found to be negligible.

\begin{figure}[htbp]
\begin{center}
\begin{overpic}[width=8.7 cm,height=3.4cm,angle=0]{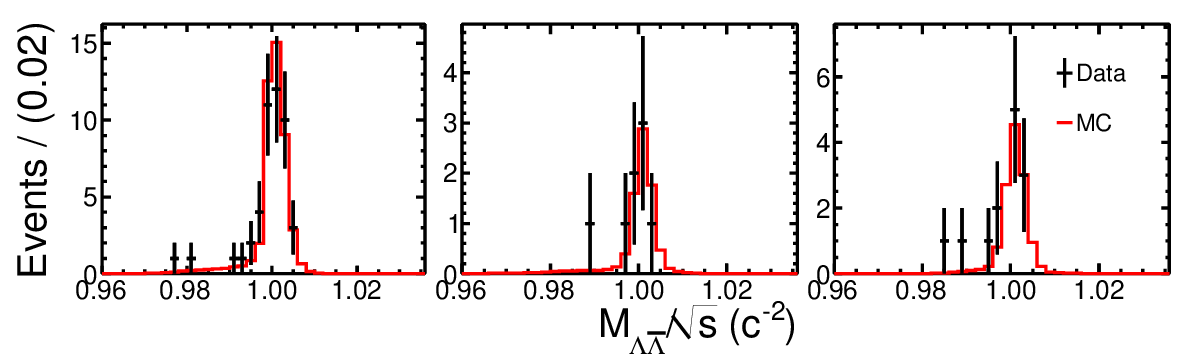}
\put(12,30){\footnotesize(a)}
\put(44,30){\footnotesize(b)}
\put(75,30){\footnotesize(c)}
\end{overpic}
\setlength{\abovecaptionskip}{-0.1cm}
\setlength{\belowcaptionskip}{-0.7cm}
\caption{Ratio of the $\lam\lambar$ invariant mass to c.m.~energy, $M_{\lam\lambar}/\sqrt{s}$, at $\sqrt{s}=$ (a) 2.400 GeV, (b) 2.800 GeV and (c) 3.080~GeV.
Dots with error bars are data
 and the solid curves (red) are the MC simulated events.}
\label{ecm}
\end{center}
\end{figure}

With the selection criteria applied, the ratios of the $\lam\lambar$ invariant
mass to c.m.~energy, $M_{\lam\lambar}/\sqrt{s}$, are shown in Fig.~\ref{ecm},
between data and signal MC.
Since the number of background events in the peaks can be neglected,
we take the number of counts in the range
of $0.98<M_{\lam\lambar}/\sqrt{s}<1.02$ as signal events, $N_{obs}$.

The Born cross sections of the process $\lamlam$
at $\sqrt{s}=2.400$, 2.800 and 3.080~GeV
are determined by Eq.~(\ref{equ:born}),
The detection efficiency $\epsilon$ is determined from  MC
The electromagnetic FF ratio, $|G_{E}/G_{M}|$, has an impact on the
angular distribution of $\lam$($\lambar$).
To address the dependency of the angular distribution of the produced baryon, the
detection efficiency is evaluated with the MC samples by sampling the baryon angular
distribution with $(1+\cos^{2}\theta)$ and $(1-\cos^{2}\theta)$,
where $\theta$ is the polar angle of $\lam$,
corresponding to the $|G_{E}|=0$ and $|G_{M}|=0$, respectively.
The nominal detection efficiency is the average of these efficiencies.

Various sources of systematic uncertainties in the cross section measurements at
$\sqs=2.400$, 2.800 and 3.080~GeV have been studied.
The uncertainty from reconstruction of $\lam(\lambar)$ and the mass window requirement is determined to be 4.5\%, as determined by a control sample of
$J/\psi\ra pK^{-}\lambar+c.c$.
The unknown angular distribution of the $\lam/\lambar$ introduces an additional
uncertainty in the efficiency. This uncertainty is estimated by taking
half of the difference between the two extremes
$(1+\cos^{2}\theta)$ and $(1-\cos^{2}\theta)$
and is within the range
 10.8\%$\sim$12.7\%, depending on the c.m.~energy.
The uncertainty from the ISR correction factor is estimated by varying the input cross section
line shape of $e^{+}e^{-}\to\Lambda\overline{\Lambda}$
within uncertainty, and is in the range 2.2\%$\sim$4.0\% depending on c.m.~energy.
The uncertainty of integrated luminosity is 1.0\%~\cite{lumi}.
The uncertainties are assumed to be uncorrelated and combined in quadrature,
giving in total of 13.0\%$\sim$14.0\% for the cross section measurements at $\sqs=2.400$, 2.800 and 3.080~GeV.

\section{\boldmath Results and conclusion}
The resulting Born cross section and
the effective FFs of $\lam$ in the timelike region, defined in
Eq.~(\ref{eq2}), at $\sqrt{s}=2.2324$, 2.400, 2.800 and 3.080~GeV are summarized in
Table~\ref{result1}.
The results at $\sqrt{s}=2.2324$~GeV from modes I and II are combined taking into account the
correlation between the uncertainties of the two decay modes~\cite{combine1, combine2}.

\begin{table}[htbp]
\setlength{\abovecaptionskip}{-0.1cm}
\setlength{\belowcaptionskip}{-0.7cm}
\caption{
The measured Born cross sections, $\sigma^{\rm B}$.
 The subscripts $1$, $2$ and $c$ denote mode I, mode II and the combined result.
 The first uncertainties are statistical and the second systematic.}
\footnotesize
\begin{center}
\begin{tabular}{c|c|c|c|c|c}
\hline
\hline
$\sqrt{s}$&   $\mathcal{L}_{\rm int}$  & $N_{\rm obs}$  & $\epsilon(1+\delta)$ & $\sigma^{\rm B}$ & $|G|$ \\
(GeV)   &   (pb$^{-1}$)  &  & (\%) &~(pb) & ($\times10^{-2}$)\\
\hline
2.2324$_{1}$ &    2.63  & $43\pm7$  & 12.9  & $312\pm51^{+72}_{-45}$\\
2.2324$_{2}$  &    2.63  & $22\pm6$ & 8.25  & $288\pm96^{+64}_{-36}$\\
2.2324$_{c}$ &    &          &             & $305\pm45^{+66}_{-36}$  & $61.9\pm4.6^{+18.1}_{-9.0}$ \\
\hline
2.400  & 3.42 &  $45\pm7$  & 25.3 & $128\pm19\pm18$     & $12.7\pm0.9\pm0.9$\\
2.800  & 3.75 &  $8\pm3$   & 36.1 & $14.8\pm5.2\pm1.9$  & $4.10\pm0.72\pm0.26$\\
3.080  & 30.73 & $13\pm4$  & 24.5 & $4.2\pm1.2\pm0.5$   & $2.29\pm0.33\pm0.14$\\
\hline
\hline
\end{tabular}
\label{result1}
\end{center}
\end{table}

\begin{figure*}[htbp]
\begin{center}
\begin{overpic}[width=8cm,height=6cm,angle=0]{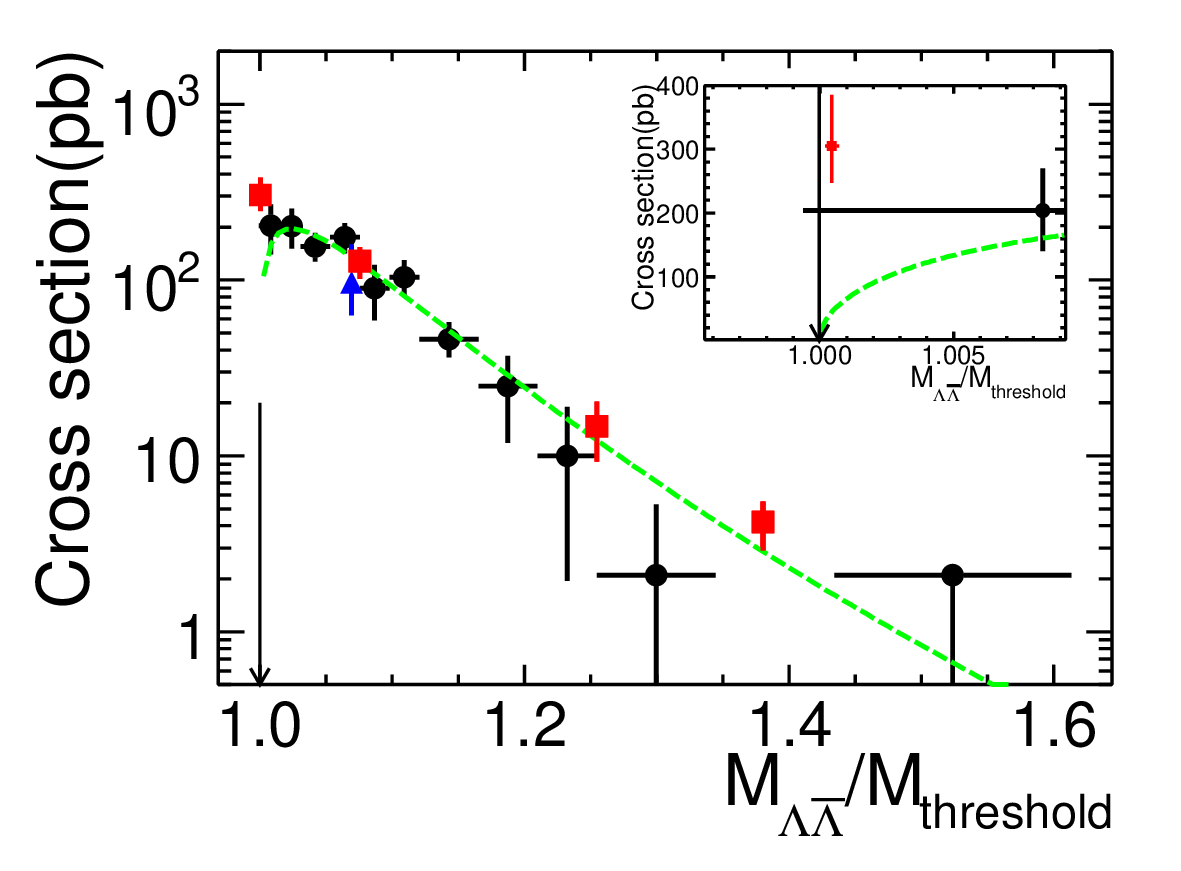}
\put(30,65){(a)}
\end{overpic}
\begin{overpic}[width=8cm,height=6cm,angle=0]{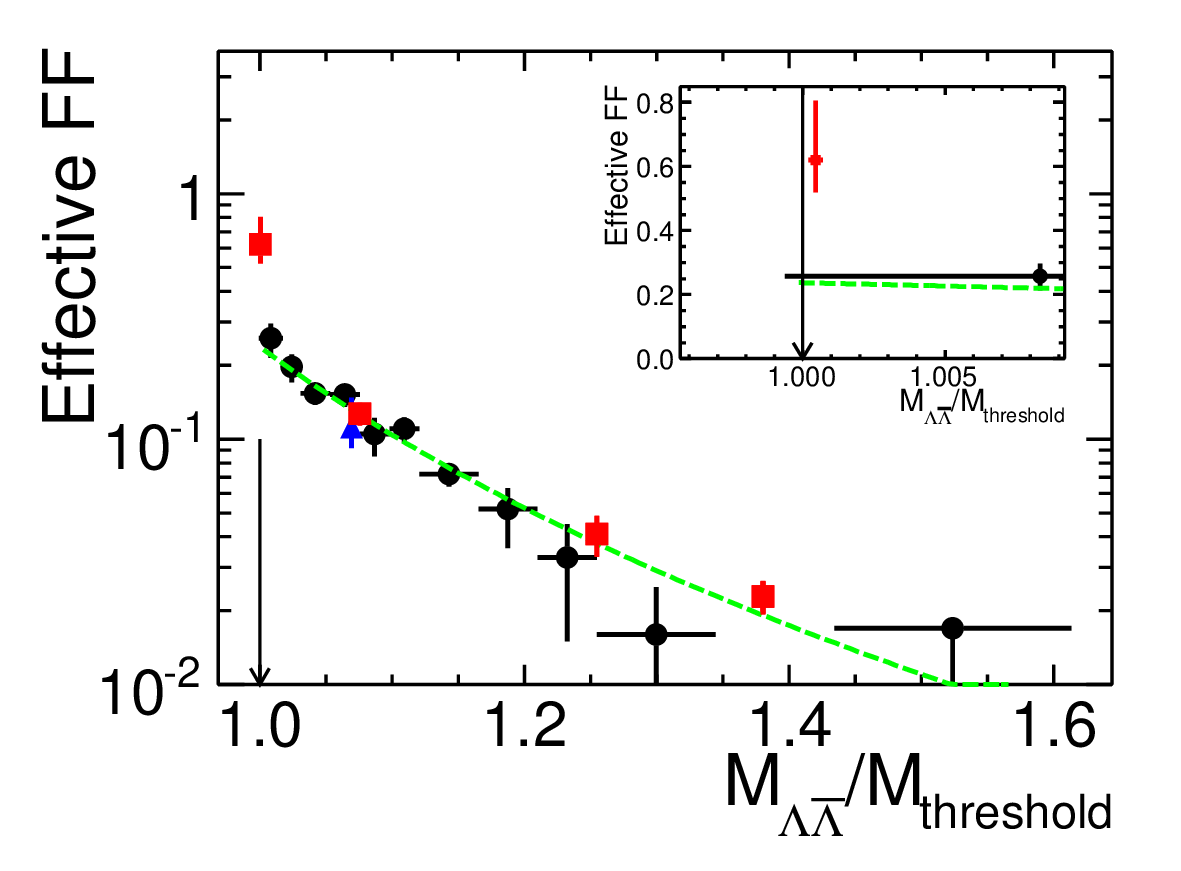}
\put(30,65){(b)}
\end{overpic}
\setlength{\abovecaptionskip}{-0.1cm}
\setlength{\belowcaptionskip}{-0.7cm}
\caption{Comparisons of (a) the Born cross section and (b) effective FF in this analysis with previous experiments for $\lam\lambar$ masses from 2.0 to 3.6~GeV/$c^{2}$.
The squared (red) represent the results from this analysis, the dots (black)
and triangles (blue) are those from BaBar and DM2 results.
The lines in dot (green) are
a phenomenological fit according to a pQCD prediction, the dotted vertical lines indicate the threshold.
The insert plots are the zoomed-in results near threshold.}
\label{compare1}
\end{center}
\end{figure*}

A comparison of the Born cross sections and the effective FFs of the process $\lamlam$
with previous experimental results is illustrated in Fig.~\ref{compare1},
with the mass of $\Lambda\overline{\Lambda}$ pair relative to the its threshold.
For better resolution at $\sqrt{s}=2.2324$~GeV, a zoomed-in results in
the near-threshold region is inserted in each plot.
Our results are consistent with previous
experiments, but with improved precision.
A phenomenological fit, according to the expectation that the cross section
should be proportional to the PHSP factor times a perturbative QCD (pQCD) driven energy power~\cite{Pacetti},
is also given in Fig.~\ref{compare1}.
The anomalous behavior differing
from the pQCD prediction at threshold is observed.

In summary, based on an integrated luminosity of 2.63~pb$^{-1}$
data collected at $\sqrt{s}=2.2324$~GeV,
which is 1.0~MeV above the $\lam\lambar$ mass threshold,
we present a measurement of the process $\lamlam$. The Born cross section at $\sqrt{s}=2.2324$~GeV is determined to be $305\pm45^{+66}_{-36}$ pb,
where the first uncertainty is statistical and the second systematic.
The result is  larger than the traditional theory expectation
for neutral baryon pairs, which predicts a vanishing cross section
at threshold according to Eq.~(\ref{equ:eq1}).
The observed threshold enhancement
implies a more complicated underlying physics scenario.
The Born cross sections of process $\lamlam$ are also measured at $\sqrt{s}=2.400$, 2.800 and 3.080~GeV,
and are in good agreement with BaBar and DM2's
results~\cite{Babar,dm2}, but with improved precision.
Furthermore, the effective electromagnetic FFs of $\lam$
are presented at each c.m.~energy.
The results in this analysis may help to understand the mechanism of baryon production
and test the theory hypotheses based on the threshold enhancement effect.

\section{\boldmath ACKNOWLEDGMENTS}
The BESIII collaboration thanks the staff of BEPCII, the IHEP computing center and the supercomputing center of USTC for their strong support. This work is supported in part by National Key Basic Research Program of China under Contract No. 2015CB856700; National Natural Science Foundation of China (NSFC) under Contracts Nos. 11235011, 11322544, 11335008, 11375205, 11425524, 11625523, 11635010, 11322544, 11375170, 11275189, 11475164, 11475169, 11605196, 11605198; the Chinese Academy of Sciences (CAS) Large-Scale Scientific Facility Program; the CAS Center for Excellence in Particle Physics (CCEPP); the Collaborative Innovation Center for Particles and Interactions (CICPI); Joint Large-Scale Scientific Facility Funds of the NSFC and CAS under Contracts Nos. U1232201, U1332201, U1532257, U1532258, U1532102; CAS under Contracts Nos. KJCX2-YW-N29, KJCX2-YW-N45; 100 Talents Program of CAS; National 1000 Talents Program of China; INPAC and Shanghai Key Laboratory for Particle Physics and Cosmology; German Research Foundation DFG under Contracts Nos. Collaborative Research Center CRC 1044, FOR 2359; Istituto Nazionale di Fisica Nucleare, Italy; Koninklijke Nederlandse Akademie van Wetenschappen (KNAW) under Contract No. 530-4CDP03; Ministry of Development of Turkey under Contract No. DPT2006K-120470; National Natural Science Foundation of China (NSFC) under Contract No. 11505010; The Swedish Resarch Council; U. S. Department of Energy under Contracts Nos. DE-FG02-05ER41374, DE-SC-0010504, DE-SC-0010118, DE-SC-0012069; U.S. National Science Foundation; University of Groningen (RuG) and the Helmholtzzentrum fuer Schwerionenforschung GmbH (GSI), Darmstadt; WCU Program of National Research Foundation of Korea under Contract No. R32-2008-000-10155-0.

\end{document}